\documentclass[twocolumn,english,showpacs,preprintnumbers,amsmath,amssymb]{revtex4}
\usepackage[T1]{fontenc}
\usepackage[latin1]{inputenc}
\usepackage{graphicx}

\makeatletter

\usepackage{dcolumn}
\usepackage{bm}

\usepackage{graphics}
\usepackage{babel}
\makeatother
\begin{document}

\title{Induced Kramer-Pesch-Effect in a Two Gap Superconductor:\\
 Application to MgB$_{2}$}

\author{A.~Gumann, S.~Graser, T.~Dahm, and N.~Schopohl}

\affiliation{Institut f\"ur Theoretische Physik, Universit\"at T\"ubingen\\
 Auf der Morgenstelle 14, D-72076 T\"ubingen, Germany}

\date{21st November 2005}

\begin{abstract}
The size of the vortex core in a clean superconductor
is strongly temperature dependent and shrinks with decreasing temperature,
decreasing to zero for $T\rightarrow0$. We study this so-called
Kramer-Pesch effect both for a single gap superconductor and
for the case of a two gap superconductor using parameters appropriate
for Magnesium Diboride.
Usually, the Kramer-Pesch effect is absent in the dirty limit.
Here, we show that the Kramer-Pesch effect exists in both bands
of a two gap superconductor
even if only one of the two bands is in the clean limit and the 
other band in the dirty limit, a case appropriate for MgB$_2$.
In this case an induced Kramer-Pesch effect appears in the dirty band.
Besides numerical results we also present
an analytical model for the spatial variation of the pairing potential
in the vicinity of the vortex center that allows a simple calculation
of the vortex core radius even in the limit $T\rightarrow0$.
\end{abstract}


\maketitle

\section{\label{sec:intro}Introduction}

In 1974 Kramer and Pesch theoretically discovered an unusual core shrinkage
of an isolated vortex for decreasing temperature which they did not
only observe in numerical solutions of Eilenberger's equations but also 
proved analytically
\cite{KramPesch}. While all other lengthscales describing the superconducting
state, especially the London penetration depth and the coherence length,
reach a saturation value with decreasing temperature, this investigation
introduced a new lengthscale in the discussion of the vortex state
of clean superconductors. It can be defined as the inverse of the slope
of the pairing potential at the vortex center and describes not only
the spatial variation of the gap function but also the maximum height
of the supercurrent density around the vortex center and thus measures the
size of the vortex core. In a clean superconductor
without impurity scattering this length $\xi_{v}$ decreases linearly
with temperature tending to zero for $T\rightarrow0$ while in superconductors
with larger scattering rates it reaches
a saturation value. A detailed study of the impurity effect on the
vortex core shrinkage can be found in Ref.~\cite{HayKatSig}. 

It appears well established both experimentally and theoretically that
the new superconductor MgB$_2$ is a two gap superconductor,
possessing two superconducting gaps of significantly different
size on disconnected parts of the Fermi surface
\cite{Kort,LiuMazKort,Choi,DahmReview}. There exist
other compounds, in which two gap superconductivity is believed
to be realized, however, at present MgB$_2$ is the clearest
example. In the case of such a multi band superconductor 
with several distinct gap values there
exist different lengthscales that describe the spatial variation of
the quasiparticle excitations corresponding to the different gap
values. Therefore, we also expect different lengthscales
for the description of the increase of the gap functions near the
center of an isolated vortex in each band. It is obvious that these lengthscales
are not independent of each other if there is a coupling of the different
bands via the pairing interaction. In this work we want to examine
numerically the vortex core shrinkage for a two gap superconductor
at the example of Magnesium Diboride. We also present an analytical
gap model that well describes the low temperature behaviour of the
core size $\xi_{v}$. In the last section we want to discuss a model
of a two gap superconductor assuming large and small scattering rates
in the different bands and its influence on the size of the vortex
core.

\section{\label{sec:singlebc}The Single Band Case}

In this section, we study the vortex core structure of a standard
(single band) superconductor. In the first subsection (\ref{sub:numcalc1}),
we describe how the Riccati parametrization formalism of the Eilenberger
theory can be used to numerically calculate the structure of an isolated
Abrikosov vortex in the clean limit. In the second subsection (\ref{sub:exsol1}),
we present an analytical model for the pairing potential in the vicinity
of an isolated Abrikosov vortex and explain how it can be used to
calculate the temperature dependence of the vortex core size. The
third subsection (\ref{sub:dirtylim1}) applies to the dirty limit.
We point out how the theoretical description simplifies in this special
case and discuss the results for the vortex core size.

\subsection{\label{sub:numcalc1}Numerical Calculations}

In order to calculate the pairing potential $\Delta(\vec{r},T)$ for
a certain region in real space, one has to find a solution to the
gap equation:\begin{eqnarray}
\Delta(\vec{r},T) & = & VN(0)\,2 \pi T\sum_{0<\varepsilon_{n}<\omega_{c}}\langle f(\vec{r},\vec{k}_{F},i\varepsilon_{n})\rangle_{FS}
\nonumber\\
\label{eq:gapeq1B}\end{eqnarray}
According to Refs.~\cite{Schop1,SchopMak}, the quasiclassical propagator
$f(\vec{r},\vec{k}_{F},i\varepsilon_{n})$ can be expressed in terms
of two complex quantities $a(x)$ and $b(x)$, called the Riccati
amplitudes: \begin{eqnarray}
f(\vec{r}(x),\vec{k}_{F},i\varepsilon_{n}) & = & \frac{2\, a(x)}{1+a(x)\, b(x)}\label{eq:propf}\end{eqnarray}
 The Riccati amplitudes $a(x)$ and $b(x)$ are in turn solutions
to the Riccati differential equations\begin{subequations}\label{eq:riccatipde}\begin{eqnarray}
\hbar v_{F}\partial_{x}\, a(x)+[2\tilde{\varepsilon}_{n}+\Delta^{\dagger}(\vec{r}(x))\, a(x)]\, a(x)-\Delta(\vec{r}(x)) & = & 0\nonumber \\
\label{eq:riccatipdea}\\\hbar v_{F}\partial_{x}\, b(x)-[2\tilde{\varepsilon}_{n}+\Delta(\vec{r}(x))\, b(x)]\, b(x)+\Delta^{\dagger}(\vec{r}(x)) & = & 0\nonumber \\
\label{eq:riccatipdeb}\end{eqnarray}
\end{subequations}The Riccati differential equations have to be solved
along real space trajectories $\vec{r}(x)$ pointing in the direction
of the Fermi wave vector $\vec{v}_{F}(\vec{k}_F)$ using the modified Matsubara
frequencies $i\tilde{\varepsilon}_{n}=i\varepsilon_{n}+(e/c)\,\vec{v}_{F}\cdot\vec{A}(\vec{r}(x))$.
The quasiclassical Green's function $f(\vec{r},\vec{k}_{F},i\varepsilon_{n})$
calculated with the Riccati amplitudes according to equation (\ref{eq:propf})
then has to be averaged over the Fermi surface of the superconducting
material (denoted with $\langle\cdots\rangle_{FS}$).

As initial values for the integration of the Riccati differential 
equations the bulk values of the Riccati
amplitudes have to be used ($\varepsilon_{n}>0$):
\begin{subequations}\label{eq:abasymp}\begin{eqnarray}
a(-\infty) & = & \frac{\Delta(-\infty)}{\varepsilon_{n}+\sqrt{\varepsilon_{n}^{2}+\left|\Delta(-\infty)\right|^{2}}}\label{eq:aasymp}\\
b(+\infty) & = & \frac{\Delta^{\dagger}(+\infty)}{\varepsilon_{n}+\sqrt{\varepsilon_{n}^{2}+|\Delta(+\infty)|^{2}}}\label{eq:basymp}\end{eqnarray}

\end{subequations}The quasiclassical Green's function $f(\vec{r},\vec{k}_{F},i\varepsilon_{n})$
is, via the Riccati amplitudes, a function of the pairing potential. A straightforward
way to find a solution to the self-consistency problem posed by the gap equation is to choose a certain
initial configuration for the pairing potential $\Delta(\vec{r},T)$
and then improve this configuration iteratively.

Since we will apply the methods described in this chapter to study
the temperature dependence of the vortex core size of an isolated
Abrikosov vortex, it is convenient to define, at this point, a characteristic
lengthscale for the size of the vortex core. We take up the proposal
for a definition by Hayashi et al. \cite{HayKatSig}:
\begin{eqnarray}
\xi_{v}^{-1} & = & \left.\frac{\partial\Delta(r)}{\partial r}\right|_{r=0}\frac{1}{\Delta(r=\infty,\, T)}\label{eq:defxiv}
\end{eqnarray}
This characteristic lengthscale $\xi_{v}$ corresponds to the inverse
slope of the pairing potential $\Delta(\vec{r},T)$ at the vortex
center $r=0$, normalized to the bulk value of the pairing potential
$\Delta(r=\infty,\, T)$ at the corresponding temperature $T$ and has to
be distinguished from the bulk coherence length
$\xi_{\infty}=\hbar v_F / \Delta(r=\infty,\, T)$.

\begin{figure}
\begin{center}
\includegraphics
[width=0.8\columnwidth,
keepaspectratio]
{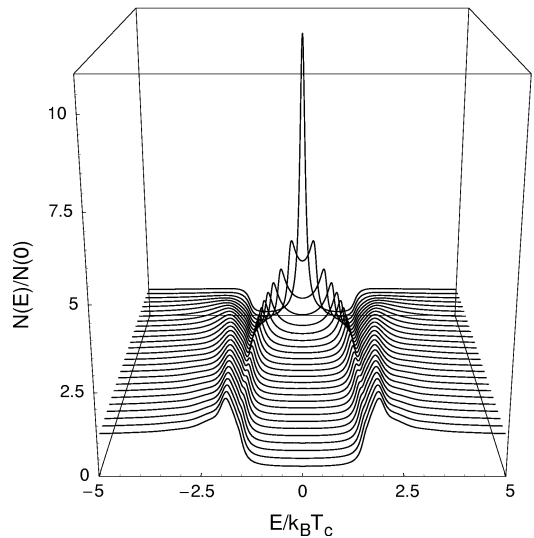}
\end{center}
\caption{\label{cap:ldos1B}The quasiparticle spectrum in the single band
case, plotted for several distances from the vortex core at a temperature
of $T=0.5\,T_c$. The distances
span from $r=0$ (rearmost curve) to $r=5\,\xi_{\infty}$(foremost
curve) in steps of $0.2\,\xi_{\infty}$. For these plots, an imaginary
part (broadening) of $\delta=0.1$ was used.}
\end{figure}

Once a self-consistent solution for the pairing potential is found,
the local quasiparticle spectrum $N(\vec{r},E)$ (local density of
states) can be calculated in the following manner:
\begin{eqnarray}
N(\vec{r},E) & = &
\left<\mbox{Re}\left[\frac{1-a\, b}{1+a\, b}\right]_{i\varepsilon_{n}\rightarrow
E+i\delta}\right>_{FS}
\label{eq:ldos1B}
\end{eqnarray}
In Fig.~\ref{cap:ldos1B}, we show numerical results for the quasiparticle
spectrum in the single band case. The curves were obtained evaluating
Eq.~(\ref{eq:ldos1B}) based on self-consistently calculated
data for the pairing potential $\Delta(\vec{r},T)$ and a cylindrical Fermi
surface has been assumed. A distinct zero energy peak represents
Andreev bound states in the vortex core. For increasing distance from 
the vortex core, this peak splits and finally merges into the gap edges.
Similar results have been obtained before. Gygi and Schl\"uter solved the 
Bogoliubov-de~Gennes equations as an eigenvalue problem to study the electronic 
structure of a vortex line, see Ref.~\cite{GySchl}. Ichioka et al.~used the 
quasiclassical Eilenberger theory to examine the same 
problem, see Ref.~\cite{Ichi}.

Although the Riccati parametrization of the Eilenberger propagator
offers a convenient and stable method to solve the Eilenberger equations and
to calculate the local density of states, the effort of numerical calculations
increases under certain circumstances. Firstly, as the lengthscale
on which the pairing potential $\Delta(\vec{r},T)$ varies is getting
shorter, the number of grid points on which the gap equation has to be
solved in real space grows. Secondly, the number of terms in the sum
over Matsubara frequencies $\sum_{0<\varepsilon_{n}<\omega_{c}}$
grows linearly with decreasing temperature. Thirdly, the averaging
over all Fermi wave vectors $\vec{v}_{F}$ that appear on the Fermi
surface becomes laborious for complicated Fermi surface topologies.

Taking all these points into account, it is obvious that an analytical
approach should be considered whenever possible, in particular
if one wants to study the $T\rightarrow0$ limit. In the next chapter,
this will be done for an isolated Abrikosov vortex.

\subsection{\label{sub:exsol1}Exact Solution for an Analytical Gap Model}

In the innermost part of an Abrikosov vortex, the modulus of the pairing
potential grows linearly with increasing distance from the vortex
center. As it approaches the bulk value, it saturates with a certain
profile. Superimposed to the behaviour of the modulus of the pairing
potential, the phase varies by $2\pi$ around the vortex. This leads
to the general form $\Delta(\vec{r})\sim r\,e^{i\phi}$ for the innermost
part of the vortex, while in the outermost part, the behaviour 
$\Delta(\vec{r})\sim e^{i\phi}$ holds ($r$ and $\phi$ are polar 
coordinates in real space). Thus, the following simple analytical model for
the pairing potential in the vicinity of an isolated Abrikosov vortex
is useful:
\begin{eqnarray}
\Delta(\vec{r},T) & = & \left\{ \begin{array}{cc}
\frac{\Delta_{\infty}(T)}{\xi_{v}}\,r\,e^{i\phi} & \mbox{,}\,r<\xi_{v}\\
\Delta_{\infty}(T)\, e^{i\phi} & \mbox{,}\,r\geq\xi_{v}\end{array}\right.
\label{eq:gaptrichter}
\end{eqnarray}
 Here, $\Delta_{\infty}(T)$ denotes the value of the pairing potential
in the bulk at a given temperature $T$; $\xi_{v}$ is the lengthscale on which
the pairing potential rises from zero in the center of the vortex to the bulk
value $\Delta_{\infty}(T)$ and coincides with the previous definition (\ref{eq:defxiv}).
According to Ref.~\cite{Schop1}, every real space trajectory $\vec{r}(x)$ is
parametrized by an impact parameter $y$ and the position along the trajectory $x$.
In our special case, the impact parameter $y$ is given by the distance from the
trajectory to the vortex core. The position along the trajectory $x$ is measured
relative to the point closest to the vortex core. It follows from what has been said that 
\begin{eqnarray}
r\,e^{i\phi}&=&(x+iy)\,e^{i\theta}
\end{eqnarray}
 along a chosen trajectory specified by $\theta$, 
the angle corresponding to the direction of the Fermi velocity in the 
$ab$-plane. Within the scope of this model, the pairing potential consists of a vortex
core with a linear profile and a phase vortex in the circumference.\\
In the next two subsections, we will show how an analytical solution
can be obtained for the quasiclassical Green's function in the special
case of this particular pairing potential profile. We will then use this
analytical solution in order to determine the lengthscale $\xi_{v}$
self-consistently with the gap equation.

\subsubsection{\label{subsub:linprof}Analytical Solution for $r<\xi_{v}$}

\noindent In general, the linearized Bogoliubov-de~Gennes equations, often referred
to as Andreev equations, read
\begin{eqnarray}
\label{BdGeq}
-i\hbar v_F\frac{\partial}{\partial x}
\left[ \begin{array}{c} u(x) \\ v(x) \end{array} \right]
& = &
\left[
\begin{array}{cc}
i\tilde{\varepsilon}_n(x) & -\Delta(x)\\
\Delta^{\dagger}(x) & -i\tilde{\varepsilon}_n(x)
\end{array}
\right]
\left[ \begin{array}{c} u(x) \\ v(x) \end{array} \right]
\end{eqnarray}
According to Ref.~\cite{Schop1}, the solutions of (\ref{BdGeq}) are connected to
the Riccati amplitude $a(x)$ via
\begin{eqnarray}
\label{BdGtoRicc}
a(x)=i\,\frac{u(x)}{v(x)}
\end{eqnarray}
whereas the Riccati amplitude $b(x)$ can be obtained applying the
symmetry relation\begin{eqnarray}
b(x) & = & -a(-x)\, e^{-2i\theta}\label{eq:absym}\end{eqnarray}
which is valid if only a single vortex line is present.\\
It should be noted that we do not solve the linearized Bogoliubov-de~Gennes 
equations~(\ref{BdGeq}) as an eigenvalue problem, but as an inital value problem, 
see Ref.~\cite{Schop1}.\\
For the linear profile, Eq.~(\ref{BdGeq}) reads
\begin{eqnarray}
 & \left\{ i\hbar v_{F}\partial_{x}+e^{i\frac{\theta}{2}\hat{\tau}_{3}}\left[i\varepsilon_{n}\hat{\tau}_{3}-\frac{\Delta_{\infty}}{\xi_{v}}i\,(x\hat{\tau}_{2}+y\hat{\tau}_{1})\right]e^{-i\frac{\theta}{2}\hat{\tau}_{3}}\right\} \nonumber \\
 & \hspace{4cm}\circ
 \left(
 \begin{array}{c}
 u(x) \\ v(x)
 \end{array}
 \right)
 \,=\,\hat{0}
\label{eq:andreev}
\end{eqnarray}
where $\hat{\tau}_i$ are the Pauli matrices. 
Applying the gauge transformation\begin{eqnarray}
\hat{\Psi}(x) 
\,=\,
\left(
\begin{array}{c}
\Psi_1(x) \\ \Psi_2(x)
\end{array}
\right)
& = & e^{-i\frac{\theta}{2}\hat{\tau}_{3}}\,\circ\,
 \left(
 \begin{array}{c}
 u(x) \\ v(x)
 \end{array}
 \right)
\label{eq:gautra}
\end{eqnarray}
 and introducing the spinor $\hat{\Phi}(x)$ via \begin{eqnarray}
\hat{\Psi}(x) & = & \left(\hbar v_{F}\partial_{x}-\varepsilon_{n}\hat{\tau}_{3}+\frac{\Delta_{\infty}}{\xi_{v}}\,(x\hat{\tau}_{2}+y\hat{\tau}_{1})\right)\nonumber \\
&& 
\hspace{3.5cm}
\circ\,e^{i\frac{\pi}{4}\hat{\tau}_{1}}\,\circ\,\hat{\Phi}(x)\label{eq:ansatzpsi}\end{eqnarray}
one obtains the following two decoupled differential equations for
the components $\hat{\Phi}(x)=(\Phi_{-}(x),\,\Phi_{+}(x))^{T}$ \begin{eqnarray}
 & \left[(\hbar v_{F}\partial_{x})^{2}-\varepsilon_{n}^{2}-\frac{\Delta_{\infty}^{2}}{\xi_{v}^{2}}\,(x^{2}+y^{2})\pm\hbar v_{F}\frac{\Delta_{\infty}}{\xi_{v}}\right]\nonumber \\
 & \hspace{4cm}\circ\,\Phi_{\mp}(x)=0\label{eq:decoup}\end{eqnarray}
Scaling the variables $\bar{x}=\frac{x}{\xi}$ and $\bar{y}=\frac{y}{\xi}$
to the characteristic lengthscale $\xi=\sqrt{\frac{\xi_{v}\,\xi_{\infty}}{2}}$
with $\xi_{\infty}=\frac{\hbar v_{F}}{\Delta_{\infty}}$ and introducing\begin{eqnarray}
\Phi_{\pm}(x) & = & \left(\frac{\xi}{\hbar v_{F}}\right)^{2}\,\bar{\Phi}_{\pm}(\bar{x})\label{eq:scaling}\\
c_{\pm} & = & \frac{1}{4}\left(\frac{\xi_{v}}{\xi}\frac{\varepsilon_{n}}{\Delta_{\infty}}\right)^{2}+\frac{1}{4}\bar{y}^{2}\pm\frac{1}{2}\label{eq:cplusminus}\end{eqnarray}
Eqs. (\ref{eq:decoup}) assume the following compact and dimensionless
form:\begin{eqnarray}
\left[\partial_{\bar{x}}^{2}-c_{\pm}-\frac{1}{4}\,\bar{x}^{2}\right]\bar{\Phi}_{\pm}(\bar{x}) & = & 0\end{eqnarray}
The general solution of this second order differential equation is given by a linear combination
of the parabolic cylinder functions $U(c_{\pm},\bar{x})$ and $V(c_{\pm},\bar{x})$
\cite{AbramStegun}. We choose the following ansatz
\begin{eqnarray}
\bar{\Phi}_{-}(\bar{x}) & = & A_{-}\, V(c_{-},-\bar{x})\,\Theta(-\bar{x})+B_{-}\, V(c_{-},\bar{x})\,\Theta(\bar{x})\nonumber \\
\bar{\Phi}_{+}(\bar{x}) & = & A_{+}\, U(c_{+},-\bar{x})\,\Theta(-\bar{x})+B_{+}\, U(c_{+},\bar{x})\,\Theta(\bar{x})\nonumber \\
\label{eq:ansatz2}\\ &  & \mbox{with}\quad\Theta(\bar{x})\,=\,\left\{ \begin{array}{c}
1\quad\mbox{for}\quad\bar{x}>0\\
0\quad\mbox{for}\quad\bar{x}<0\end{array}\right.\nonumber
\end{eqnarray}
in order to construct the Riccati amplitudes $a(x)$ and $b(x)$.
The connection between the Riccati amplitude $a(x)$ and the solutions
of the Andreev equation (\ref{eq:andreev}), given by (\ref{BdGtoRicc}), now reads
\begin{eqnarray}
\label{BdGtoRicc2}
a(x) & = & \frac{\Psi_{1}(x)}{\Psi_{2}(x)}\, ie^{i\theta}\label{eq:}
\end{eqnarray}
Let us briefly comment on the construction of the Riccati amplitudes
$a(x)$ and $b(x)$, respectively. A correct solution firstly has to
solve the Riccati equations (\ref{eq:riccatipde}) and secondly has
to possess the correct bulk asymptotics (\ref{eq:abasymp}). The former
is provided by construction \cite{Schop1} and we will see later on
that the asymptotic behaviour obtained with this ansatz is indeed
correct.\\
Additionally, it should be noted that the ansatz (\ref{eq:ansatz2})
is in fact unique. An expansion of the solution derived from the most
general ansatz (containing both parabolic cylinder functions
$U(c_{\pm},\bar{x})$ and $V(c_{\pm},\bar{x})$ for both functions $\bar{\Phi}_{\pm}$
and all $x$) exhibits the correct asymptotic behaviour if and
only if the terms not contained in (\ref{eq:ansatz2}) vanish. Thus, the above form is
being reproduced.

Using the ansatz (\ref{eq:ansatz2}), undoing the gauge transformation
(\ref{eq:gautra}) and the substitutions (\ref{eq:ansatzpsi}), (\ref{eq:scaling}),
then applying some basic properties of the parabolic cylinder functions
\cite{AbramStegun} and making use of (\ref{BdGtoRicc2}),
one obtains the following expression for the
Riccati amplitude $a(\bar{x})$ for $\bar{x}<0$:
\begin{widetext}
\begin{eqnarray}
a(\bar{x}) & = & e^{-i\theta}\,\frac{\left[-\frac{V(c_{+},-\bar{x})}{U(c_{-},-\bar{x})}-\frac{1}{2}\left(\frac{\xi_{v}}{\xi}\frac{\varepsilon_{n}}{\Delta_{\infty}}-i\bar{y}\right)\frac{V(c_{-},-\bar{x})}{U(c_{-},-\bar{x})}\right]-\left(\frac{A_{+}}{iA_{-}}\right)\left[1-\frac{1}{2}\left(\frac{\xi_{v}}{\xi}\frac{\varepsilon_{n}}{\Delta_{\infty}}+i\bar{y}\right)\frac{U(c_{+},-\bar{x})}{U(c_{-},-\bar{x})}\right]}{\left[-\frac{V(c_{+},-\bar{x})}{U(c_{-},-\bar{x})}+\frac{1}{2}\left(\frac{\xi_{v}}{\xi}\frac{\varepsilon_{n}}{\Delta_{\infty}}-i\bar{y}\right)\frac{V(c_{-},-\bar{x})}{U(c_{-},-\bar{x})}\right]+\left(\frac{A_{+}}{iA_{-}}\right)\left[1+\frac{1}{2}\left(\frac{\xi_{v}}{\xi}\frac{\varepsilon_{n}}{\Delta_{\infty}}+i\bar{y}\right)\frac{U(c_{+},-\bar{x})}{U(c_{-},-\bar{x})}\right]}
\label{eq:aofx}
\end{eqnarray}
\end{widetext}

A similar expression can be obtained for $\bar{x}>0$, but this will
not be necessary for our further calculations. The Riccati amplitude
$b(\bar{x})$ for $\bar{x}>0$ can be constructed from equation (\ref{eq:aofx}) using the
symmetry relation (\ref{eq:absym}). We now have to determine the
coefficient $(\frac{A_{+}}{iA_{-}})$ in order to complete the solution.
This will be achieved by matching Eq. (\ref{eq:aofx}) to the solution
for $r\geq\xi_{v}$.

\subsubsection{\label{subsub:phasevortex}Analytical Solution for $r\geq\xi_{v}$}

The outer region of our model for the pairing potential (\ref{eq:gaptrichter})
is nothing but a pure phase vortex. Unfortunately, there is no exact
solution to the Riccati equations (\ref{eq:riccatipde}) for such a phase
vortex for all energies and all impact parameters. On the one hand,
an exact solution for all values of the impact parameter can be obtained
for the special case $i\varepsilon_{n}\rightarrow E=|\Delta_{\infty}|$
\cite{Schop1}. On the other hand, an asymptotic solution for all energies
is obvious for vanishing impact parameter (In this special case, the pairing
potential is constant along any single trajectory except for the phase step in the
vortex core). Since we will solve the gap equation for a real space position $x_{0}$ that is very
close to the vortex center (in fact we will perform the limit $x_{0}\rightarrow0$),
the impact parameter will be vanishingly small in our calculations.
Thus, we choose the asymptotic solution that is exact for all energies and vanishing
impact parameter. It is of the following form:
\begin{eqnarray}
a(\bar{x}) & = & \frac{|\Delta_{\infty}|}{\varepsilon_{n}+\sqrt{\varepsilon_{n}^{2}+|\Delta_{\infty}|^{2}}}\, e^{i\phi}
\label{eq:aofxBulk}
\end{eqnarray}
Now the solution for the outer part of the vortex being constructed,
we can demand that the solution be continuous at $\xi_{v}$. This means
that Eq.~(\ref{eq:aofx}) evaluated at $x=\xi_v$ has to be equal to
Eq.~(\ref{eq:aofxBulk}) which
determines the coefficient $(\frac{A_{+}}{iA_{-}})$. Hence, we found
a solution for the Riccati amplitudes for all positions $x$ and all
energies which is exact for vanishing impact parameter $y$.

We still have to check if the asymptotic behaviour of our solution
is consistent with the bulk asymptotics of the Riccati
amplitudes. These have been discussed in section~(\ref{sub:numcalc1})
and are given in Eqs.~(\ref{eq:abasymp}). On the one hand, we insert
the linear profile of the pairing potential into Eqs. (\ref{eq:abasymp})
and thus obtain the correct asymptotic behaviour. On the other hand,
we move the matching point of our analytic solution to infinity and
then expand Eq.~(\ref{eq:aofx}) in terms of $\bar{x}\rightarrow\pm\infty$.
Indeed, this reproduces the correct bulk asymptotics. With a finite
matching point, particularly the matching point $\xi_{v}$, the correct
asymptotic behaviour is provided by construction of the solution.

Altogether, we can therefore state that the solution constructed with
ansatz (\ref{eq:ansatz2}) indeed exhibits the correct asymptotic
behaviour in the bulk.

\subsubsection{\label{subsub:gapeq}Solving the Gap Equation}

In order to determine the lengthscale $\xi_{v}$ on which the pairing
potential rises from zero to the bulk value, we have to solve the
gap equation (\ref{eq:gapeq1B}) self-consistently for a real space position $x_0$
which is close to the vortex center. Therefore, we have to calculate
the quasiclassical Green's function $f(\vec{r},\vec{k}_{F},i\varepsilon_{n})$
according to Eq. (\ref{eq:propf}) using the solutions for the Riccati
amplitudes $a(x)$ and $b(x)$ we found in subsections (\ref{subsub:linprof})
and (\ref{subsub:phasevortex}). Since we are only interested in the
lengthscale $\xi_{v}$ which corresponds to the inverse slope of the
pairing potential in the vortex core (see Eq. (\ref{eq:defxiv})),
we differentiate the gap equation (\ref{eq:gapeq1B}) at the vortex
center $x_{0}=0$ with respect to the position~$x_{0}$:\begin{eqnarray}
\frac{\Delta_{\infty}(T)}{\xi_{v}} & = & VN(0)\,2\pi T\sum_{0<\varepsilon_{n}<\omega_{c}}\langle\partial_{x_{0}}f(x_{0}=0)\rangle_{FS}\nonumber \\
\label{eq:gapeq1Bderiv}\end{eqnarray}
We now see that we only need to calculate the derivative of the Green's
function $f(\vec{r},\vec{k}_{F},i\varepsilon_{n})$ at the vortex
center:\begin{eqnarray}
\label{eq:dfx0}
\partial_{x_{0}}f(x_{0}=0) & = & \left.\frac{\partial}{\partial x_{0}}\frac{2\, a}{1+a\, b}\right|_{x_{0}=0}\end{eqnarray}
We therefore only need to know the quantities 
$a(x_{0}=0)$ and $\partial_{x_{0}}a(x_{0}=0)$
which correspond to a vanishing impact parameter $y$. The Riccati
amplitude $b(x)$ can then be calculated using the symmetry relation
(\ref{eq:absym}). The expression obtained this way then has to be
averaged over the Fermi surface (denoted by $\langle\cdots\rangle_{FS}$) and
inserted into the gap equation. The lengthscale $\xi_{v}(T)$
is, for the full temperature range, determined by this equation. Note
that $\xi_v$ still has to be determined self-consistently, because the Riccati
amplitude $a(x)$ depends on $\xi_v$ via Eq.~(\ref{eq:aofx}).

\begin{figure}
\begin{center}\includegraphics[%
  width=0.95\columnwidth,
  keepaspectratio]{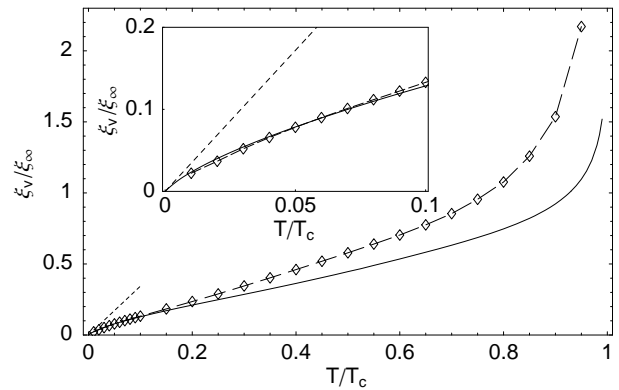}\end{center}

\caption{\label{cap:res1B}The lengthscale $\xi_{v}(T)$ obtained from a linear
profile of the pairing potential (solid line) and from a numerical
self-consistent solution of the gap equation (diamonds with dashed
guideline). Additionally, the plot shows the inclination of $\xi_{v}$
in the limit $T\rightarrow0$ (dotted line). The inset is a blow-up of the low
temperature region.}
\end{figure}

\begin{figure}
\begin{center}\includegraphics[%
  width=0.95\columnwidth,
  keepaspectratio]{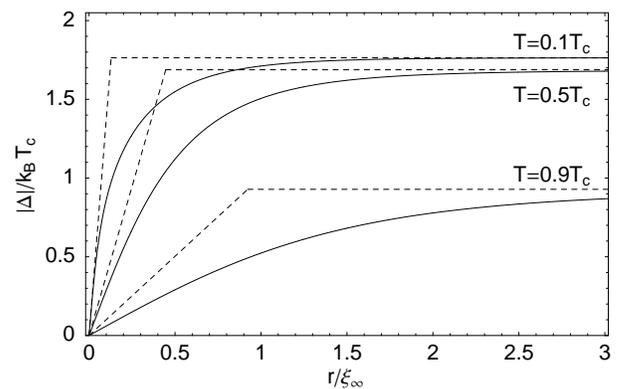}\end{center}

\caption{\label{cap:Delta1B}The modulus of the pairing potential
as a function of the distance from the vortex center for three different temperatures. For
each of the three different temperatures, we show the numerically determined
self-consistent solution (solid line) and the linear profile with a slope calculated evaluating the
gap equation at a position $x_0 \rightarrow 0$ (dashed line).
The matching of the slope at $r=0$ becomes very good at low temperatures.}
\end{figure}

Fig.~\ref{cap:res1B} shows the results from these calculations compared
to those from a fully numerical self-consistent solution of the gap
equation described in section (\ref{sub:numcalc1}). In both cases, a
cylindrical Fermi surface has been used. The model for
the pairing potential (\ref{eq:gaptrichter}) yields a qualitatively
correct behaviour of $\xi_{v}(T)$ over the full temperature range.
The lengthscale $\xi_{v}$ diverges for $T\rightarrow T_{c}$ and
vanishes for $T\rightarrow0$. In Fig.~\ref{cap:Delta1B} we show the
numerically determined gap profile together with the approximate one using
the model given in Eq.~(\ref{eq:gaptrichter}). At low temperatures the slope at the vortex
center is reproduced very well by the approximate model, while at higher
temperatures deviations occur. This is reflected in the behaviour of the
lengthscale $\xi_v$ in Fig.~\ref{cap:res1B}.

Based on the analytical expressions obtained in this chapter, it is
possible to calculate $\xi_{v}$ in the limit $T\rightarrow0$ as
long as the cut-off frequency $\omega_{c}$ is finite. Therefore, one has to 
insert the analytical result for the Riccati amplitude $a(x)$ given in 
Eq.~(\ref{eq:aofx}) into Eq.~(\ref{eq:dfx0}) making use of the symmetry relation 
(\ref{eq:absym}); this quantity can then be inserted into the differentiated
gap equation (\ref{eq:gapeq1Bderiv}) and the limit $T\rightarrow 0$ can be
performed. The result is the following:
\begin{eqnarray}
\xi_{v} & \rightarrow & \frac{4\,\xi_{\infty}}{\pi\, VN(0)\,\Delta_{\infty}}\,\frac{T}{T_{c}}\quad\mbox{for}\quad T\rightarrow0
\end{eqnarray}
This leads to $\xi_{v}\propto\xi_{\infty}T/T_{c}$ and we can thus,
within the Riccati parametrization formalism of the Eilenberger theory,
reproduce the result found earlier by Kramer and Pesch \cite{KramPesch}
as a limiting case of our calculations.

It should be noted that the linear behaviour for $T\rightarrow0$
is valid only for a very narrow temperature range (see the inset
in Fig.~\ref{cap:res1B}). The curve $\xi_{v}(T)$ bends
towards $\xi_{v}=0$ not until very low temperatures ($\sim0.05\, T_{c}$).
Thus, linear extrapolations from higher temperatures towards $T=0$
should be handled with care.

\subsection{\label{sub:dirtylim1}Comparison with Dirty Limit Calculations}

\begin{figure}
\begin{center}\includegraphics[%
  width=0.95\columnwidth,
  keepaspectratio]{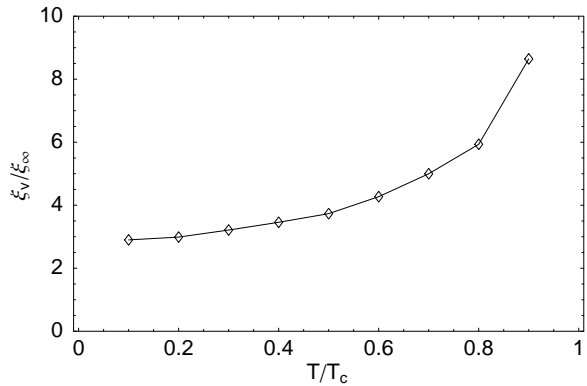}\end{center}

\caption{\label{cap:res1Busadel}The lengthscale $\xi_{v}(T)$ for the dirty
limit approach. A saturation occurs at low temperatures.}
\end{figure}

In the case of a dirty superconductor with high scattering rates the
Eilenberger equations reduce to a simple diffusionlike equation for
the Usadel propagator represented by the Green's functions $G(\vec{r},\varepsilon_{n})$
and $F(\vec{r},\varepsilon_{n})$ that are independent of the Fermi wave
vector $\vec{k}_{F}$ as was shown by Usadel \cite{Usadel}. Following
the notation of Koshelev and Golubov in \cite{Koshelev} this equation
can be written as 
\[
\varepsilon_{n}F-\frac{\mathcal{D}}{2}\left[G\left(\vec{\nabla}-\frac{2\pi i}{\Phi_{0}}
\vec{A}\right)^{2}F-F\vec{\nabla}^{2}G\right]=\Delta G
\]
together with the self-consistency condition
\begin{equation}
\Delta(\vec{r})=VN(0)\, 2\pi T \sum_{0<\varepsilon_{n}<\omega_{c}} F(\vec{r},\varepsilon_{n})
\label{eq:Gap_Usadel}
\end{equation}
where we have assumed that the Fermi surface is rotationally
symmetric with respect to the applied magnetic field and we have reduced
the equations to the single band case. The characteristic lengthscale
is determined by the diffusion constant $\mathcal{D}$ and can be
written as 
\begin{equation}
\xi_{\infty}=\sqrt{\frac{\mathcal{D}}{2\pi T_{c}}}\label{eq:coherence_length_Usadel}
\end{equation}
Employing the normalization condition
\begin{eqnarray*}
\left[G(\vec{r},\varepsilon_{n})\right]^{2}+F^{\dagger}
(\vec{r},\varepsilon_{n})F(\vec{r},\varepsilon_{n})
&=&1
\end{eqnarray*}
we can introduce a parametrization for the 
momentum averaged Usadel Green's functions $G(\vec{r},\varepsilon_{n})$ and $F(\vec{r},\varepsilon_{n})$ similar 
to the Riccati parametrization for the clean limit Eilenberger equations \cite{VavilovLarkin,EschrigKopu}. 
\begin{equation}
G(\vec{r},\varepsilon_{n})=\frac{1-a^2}{1+a^2},
\,\,\, 
F(\vec{r},\varepsilon_{n}) e^{-i\varphi}=\frac{2a}{1+a^2}
\label{eq:riccatiusadel}
\end{equation}
It should be noted that here we restricted to positive Matsubara frequencies $\varepsilon_n>0$.\\
In the vicinity of an Abrikosov vortex we can simplify the equations by 
switching to cylindrical coordinates. If we assume a large 
Ginzburg-Landau parameter $\kappa\gg1$ and confine
to an isolated vortex we can neglect the vector potential and write
down the differential equation for $a(r)$ as
\begin{eqnarray}
& & 2 \varepsilon_n a - {\mathcal{D}} \left[ \frac{1}{r}\partial_r(r\partial_r a) - \frac{2a \left(\partial_r a \right)^2}{1+a^2}
-\frac{1}{r^2} \frac{a \left(1-a^2\right)}{1+a^2} \right] \nonumber \\
& & \hspace{4.5cm} = (1-a^2)\Delta(r)
\label{diffriccatiusadel}
\end{eqnarray} 
Here $r$ denotes the distance from the vortex center and the phase of the order parameter $\varphi$ has been assumed to coincide with the polar angle $\phi$ as $\Delta(\vec{r})=\Delta(r)e^{i\phi}$.
The parameter $a(r)$ is related to the parameter $\theta$ of \cite{Koshelev} by
$a(r)=\tan\frac{\theta}{2}$. Inserting this relation into Eq.~(\ref{diffriccatiusadel}) and normalizing
all energies to $\pi T_c$ and all lengths to $\xi_\infty$ we reproduce the
differential equation for $\theta(r)$ from Ref.~\cite{Koshelev}:
\begin{eqnarray}
 &  & \partial_{r}^{2}\theta(r,\varepsilon_{n})+\frac{1}{r}\partial_{r}\theta(r,\varepsilon_{n})-\frac{1}{r^{2}}\cos\theta(r,\varepsilon_{n})\sin\theta(r,\varepsilon_{n})\nonumber \\
 &  &  +\Delta(r)\cos\theta(r,\varepsilon_{n})-\varepsilon_{n}\sin\theta(r,\varepsilon_{n})=0\label{eq:diff_equation_theta}
\end{eqnarray}
together with the gap equation 
\begin{equation}
\Delta(r)=VN(0)\,2\pi T\sum_{0<\varepsilon_{n}<\omega_{c}}\sin\theta(r,\varepsilon_{n})\label{eq:gap_equation_theta}
\end{equation}
The boundary conditions for the differential equation (\ref{eq:diff_equation_theta})
follow from general considerations: In the vortex center the anomalous
Green's function $F(\vec{r},\varepsilon_{n})$ vanishes together with
the modulus of the pairing potential leading to 
\begin{equation}
\theta(r=0,\varepsilon_{n})=0\label{eq:boundary_1}
\end{equation}
and far from the vortex center we expect the Green's functions to assume
their bulk values and we obtain
\begin{equation}
\theta(r\rightarrow\infty,\varepsilon_{n})=\arctan\frac{\Delta_{0}}{\varepsilon_{n}}\label{eq:boundary_2}
\end{equation}
To obtain a self-consistent profile for the pairing potential $\Delta(r)$
the Usadel equations have to be solved iteratively together with the
gap equation as described in section (\ref{sub:numcalc1}). We used
a relaxation method for a fast and stable solution of the boundary
value problem defined by Eqs. (\ref{eq:diff_equation_theta}, \ref{eq:boundary_1},
\ref{eq:boundary_2}) and chose a cut-off for the Matsubara frequency
summation in Eq. (\ref{eq:gap_equation_theta}) of $\omega_{c}=20\pi T_{c}$.
Extracting the characteristic length $\xi_{v}$ from the inverse slope
of the pairing potential in the vortex center we find -- as expected
from several other calculations, e.g. \cite{HayKatSig} -- a saturation
value of $\xi_{v}$ for low temperatures, see Fig.~\ref{cap:res1Busadel},
i.e. in the dirty limit the Kramer-Pesch effect disappears.

\begin{figure}
\begin{center}\includegraphics[%
  width=0.8\columnwidth,
  keepaspectratio]{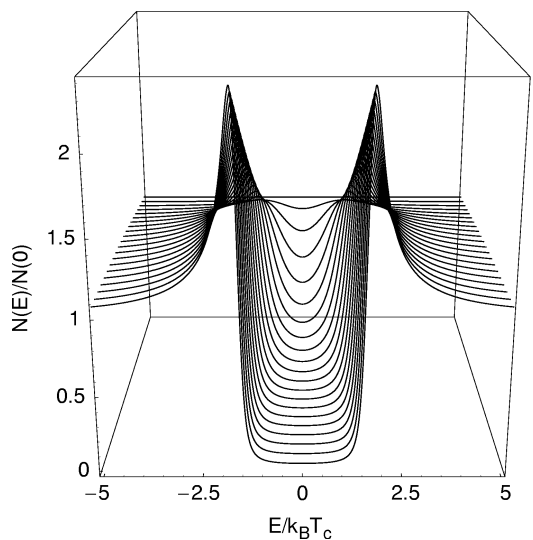}\end{center}

\caption{\label{cap:ldos1Busadel}The quasiparticle spectrum in the dirty
limit, plotted for several distances from the vortex core at a temperature
of $T=0.5\,T_c$. The distances
span from $r=0$ (rearmost curve) to $r=10\,\xi_{\infty}$ (foremost
curve) in steps of $0.5 \xi_{\infty}$.}
\end{figure}

In Fig.~\ref{cap:ldos1Busadel}, we show the quasiparticle spectrum
calculated as the real part of the analytical continuation of the
Usadel Green's function $G(\vec{r},-iE+\delta)$ according to Eq.~(\ref{eq:diff_equation_theta}).
Due to strong impurity scattering, the zero energy bound state in 
the vortex core is replaced by a flat normal state spectrum.

\section{\label{sec:twobc}The Two Gap Case}

In this section, we study the vortex core structure of a two gap
superconductor. In the first subsection (\ref{sub:numcalc2}), we
explain how the numerical procedure for the single band case from
section (\ref{sub:numcalc1}) can be generalized in order to describe
a clean two band system. In the second subsection (\ref{sub:exsol2}), we
do the same for the analytical model for the pairing potential from
section (\ref{sub:exsol1}). In this subsection, we also present results
for a two gap superconductor in the dirty limit.

\subsection{\label{sub:numcalc2}Numerical Calculations}

In order to include two band behaviour into our considerations, the
gap equation (\ref{eq:gapeq1B}) has to be replaced by a two band
gap equation:
\begin{eqnarray}
\label{eq:gapeq2B}
\Delta^{(\alpha)}(\vec{r},T)
&=&
2\pi T\,
\sum_{\alpha'}
\lambda_{\alpha \alpha'}
\sum_{0<\varepsilon_n<\omega_c}
\langle f_{\alpha'}\rangle_{FS,\alpha'}
\\
&&\hspace{1.5cm}\mbox{with}\quad
\alpha,\alpha'\in\{\sigma,\pi\}
\nonumber
\end{eqnarray}
This two band gap equation can be derived starting from the fully
momentum dependent multiband formulation of the quasiclassical (Eilenberger)
theory \cite{RieScharnSchop} and has been used successfully to describe
the upper critical field in MgB$_{2}$ \cite{DahmSchop}. We introduced two
band indices ($\sigma$ and $\pi$) with respect to the usual nomenclature
in MgB$_{2}$ as well as a $2\times 2$ coupling matrix $\lambda_{\alpha\alpha'}$.

The application of the two band gap equation (\ref{eq:gapeq2B}) requires
the calculation of the quasiclassical Green's function
for the two bands separately, $f_{\sigma}(\vec{r},\vec{k}_{F},i\varepsilon_{n})$
for the $\sigma$ band and $f_{\pi}(\vec{r},\vec{k}_{F},i\varepsilon_{n})$
for the $\pi$ band. Eqs.~(\ref{eq:propf})-(\ref{eq:abasymp})
still apply, but the $\sigma$ component of the pairing potential
$\Delta^{(\sigma)}(\vec{r},T)$ has to be used for the calculation
of $f_{\sigma}(\vec{r},\vec{k}_{F},i\varepsilon_{n})$ and 
the $\pi$ component $\Delta^{(\pi)}(\vec{r},T)$ for the calculation
of $f_{\pi}(\vec{r},\vec{k}_{F},i\varepsilon_{n})$, respectively.

Additionally, the Fermi surface topology of MgB$_{2}$ has to be taken into
account in the Fermi surface averages $\langle\cdots\rangle_{FS,\sigma}$ and $\langle\cdots\rangle_{FS,\pi}$,
respectively. Band structure calculations have been carried out to
determine the Fermi surface structure \cite{Shul,Kort} as well as
microscopic calculations that revealed the distribution of the superconducting
gap on the Fermi surface \cite{LiuMazKort,Choi}. Here, we use a simple model for
the Fermi surface that has been suggested in Ref.~\cite{DahmSchop}:
a half-torus for the $\pi$ band and a distorted cylinder for the 
$\sigma$ band, respectively. This model correctly reproduces the Fermi
surface topology and has proved to describe the upper critical field 
anisotropy in MgB$_{2}$. 
This leads to the following integral parametrizations
\cite{DahmSchop,GrasDahmSchop}:\begin{eqnarray}
\langle\cdots\rangle_{FS,\sigma} & = & \frac{c}{4\pi^{2}}\int_{-\frac{\pi}{2}}^{\frac{\pi}{2}}dk_{c}\int_{0}^{2\pi}d\phi\,\cdots\label{eq:}
\\
\langle\cdots\rangle_{FS,\pi} & = & \frac{1}{2\pi}\int_{\frac{\pi}{2}}^{\frac{3\pi}{2}}d\theta\int_{0}^{2\pi}d\phi\,\frac{1+\kappa\cos\theta}{1-2\kappa/\pi}\,\cdots\nonumber 
\nonumber
\label{eq:}\end{eqnarray}
Here, $\phi\in[0,2\pi]$ is the azimuthal angle within the \textit{ab} plane,
$\theta\in[\frac{\pi}{2},\frac{3\pi}{2}]$ the polar angle of the
torus, $k_{c}$ the \textit{c}-axis component of the momentum, $c$
the lattice constant in the \textit{c}-axis direction, and $\kappa=0.25$
the ratio of the two radii of the torus \cite{DahmSchop}.

In the two band gap equation (\ref{eq:gapeq2B}), the coupling between
the two bands occurs via the coupling matrix $\lambda_{\alpha\alpha'}$
which, in the weak coupling limit, can be calculated from the ratio
of the gap amplitudes for $T\rightarrow0$ ($\Delta_{\infty}^{(\sigma)}/\Delta_{\infty}^{(\pi)}(T\rightarrow 0)$)
and for $T\rightarrow T_{c}$ ($\Delta_{\infty}^{(\sigma)}/\Delta_{\infty}^{(\pi)}(T\rightarrow T_{c})$),
from the ratio of the densities of states in the normal (non-superconducting)
state ($N(0)^{(\sigma)}/N(0)^{(\pi)}$) and from the cut-off frequency
$\omega_{c}$ \cite{Inio}. The ratio $\Delta_{\infty}^{(\sigma)}/\Delta_{\infty}^{(\pi)}$
can be extracted from calculations of the two superconducting energy gaps based on
the Eliashberg formalism as well as the ratio $N(0)^{(\sigma)}/N(0)^{(\pi)}$
\cite{Choi}. The values $\Delta_{\infty}^{(\sigma)}/\Delta_{\infty}^{(\pi)}(T\rightarrow 0)=3$,
$\Delta_{\infty}^{(\sigma)}/\Delta_{\infty}^{(\pi)}(T\rightarrow T_{c})=4$ and
$N(0)^{(\sigma)}/N(0)^{(\pi)}=0.734$ have been used in our calculations.\\
The general form of the two band gap equation (\ref{eq:gapeq2B})
leads to the complication that the two components of the pairing potential,
$\Delta^{(\sigma)}(\vec{r,}T)$ and $\Delta^{(\pi)}(\vec{r,}T)$,
have to be calculated simultaneously.

\begin{figure}
\begin{center}
\includegraphics[%
  width=0.78\columnwidth,
  keepaspectratio]{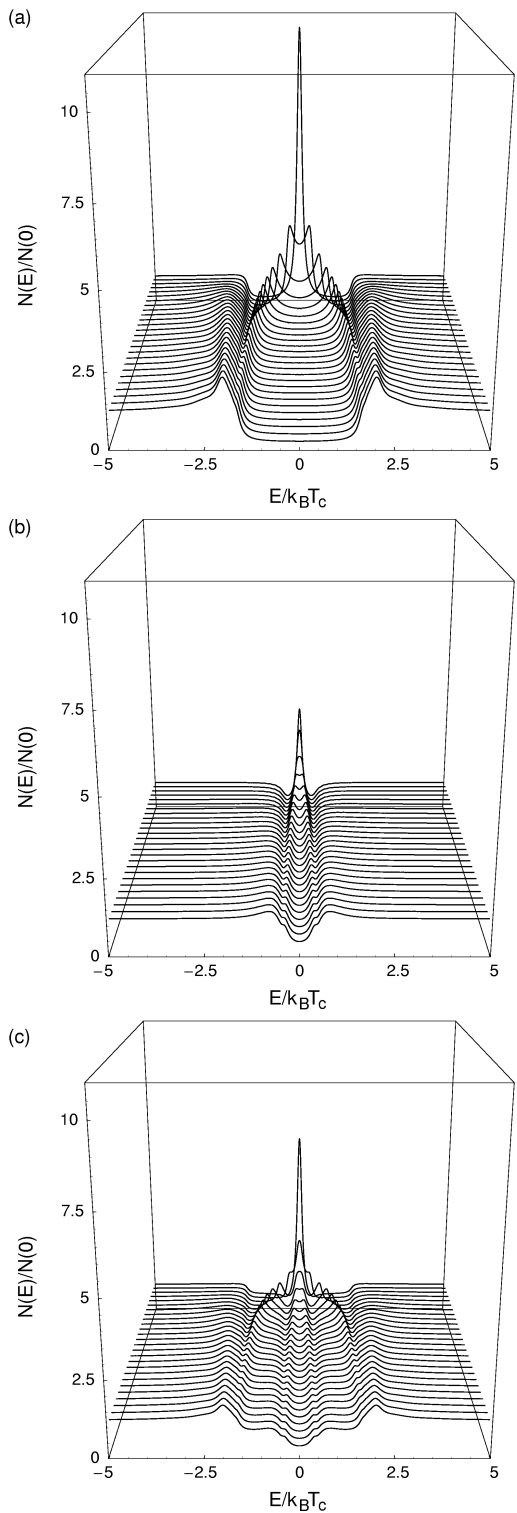}\end{center}

\caption{\label{cap:ldos2B}The quasiparticle spectrum in the two band case,
again for several distances from the vortex core. (a) shows
the quasiparticle spectrum of the $\sigma$ band and (b) that of the
$\pi$ band. In (c), the total quasiparticle spectrum is being plotted.
As before, the distances span from $r=0$ (rearmost curve) to $r=5\,\xi_{\infty}$(foremost
curve) in steps of $0.2\,\xi_{\infty}$ and an imaginary part of $\delta=0.1$
was used.}
\end{figure}

In the case of a two band superconductor, the spectral distribution
of quasiparticles $N(\vec{r},E)$ is the weighted sum of the spectral
distribution for the $\sigma$ band and for the $\pi$ band, each
of the two calculated according to Eq. (\ref{eq:ldos1B}):
\begin{eqnarray}
N_{tot}(\vec{r},T) & = & w_{\sigma}\, N_{\sigma}(\vec{r},E)+w_{\pi}\, N_{\pi}(\vec{r},E)
\end{eqnarray}
For the special case of MgB$_{2}$, the weight of the spectral distribution
of the $\pi$ band $w_{\pi}$ is given by
$w_{\pi}=N(0)^{(\pi)}/(N(0)^{(\sigma)}+N(0)^{(\pi)})=0.577$, which implies 
$w_{\sigma}=1-w_{\pi}=0.423$ (e.g. see \cite{Kort,LiuMazKort}).\\
In Fig.~\ref{cap:ldos2B}, we show numerical results for the two
band quasiparticle spectrum in the special case of MgB$_{2}$. The
total quasiparticle spectrum exhibits much more structure due to the
superposition of the spectra of the two single bands.

\subsection{\label{sub:exsol2}The analytical Gap Model for Two Bands}

In order to extend the analytical gap model from chapter (\ref{sub:exsol1})
to the case of a two gap superconductor, one has to introduce two
separate gap models for the pairing potential, one for each band,
and each of the two in the form of Eq.~(\ref{eq:gaptrichter}). Hence,
one has to introduce two lengthscales, one for each of the two bands:
$\xi_{v}^{(\sigma)}$ and $\xi_{v}^{(\pi)}$. Following the discussion
in section (\ref{sub:numcalc2}), the calculations carried out in
chapter (\ref{sub:exsol1}) have then to be repeated for each of the
two bands separately in order to obtain the quasiclassical Green's function for
both bands: $f_{\sigma}(\vec{r},\vec{k}_{F},i\varepsilon_{n})$
and $f_{\pi}(\vec{r},\vec{k}_{F},i\varepsilon_{n})$. A coupling of
the two bands first occurs in the gap equation (\ref{eq:gapeq2B}).

In full analogy to the procedure in chapter (\ref{sub:exsol1}), the
Green's function $f(\vec{r},\vec{k}_{F},i\varepsilon_{n})$ can be
calculated for each of the two bands. The two band gap equation (\ref{eq:gapeq2B})
can again be differentiated at the center of the vortex $x_{0}=0$
with respect to the position $x_{0}$ and determines the two lengthscales
$\xi_{v}^{(\sigma)}$ and $\xi_{v}^{(\pi)}$ for the full temperature
range:
\begin{eqnarray}
\frac{\Delta_{\infty}^{(\alpha)}}{\xi_{v}^{(\alpha)}}
&=&
2\pi T\,
\sum_{\alpha'}
\lambda_{\alpha \alpha'}
\sum_{0<\varepsilon_n<\omega_c}
\langle \partial_{x_0}f_{\alpha'}(x_0=0)\rangle_{FS,\alpha'}
\nonumber
\\
\end{eqnarray}

\begin{figure}
\begin{center}\includegraphics[%
  width=0.95\columnwidth,
  keepaspectratio]{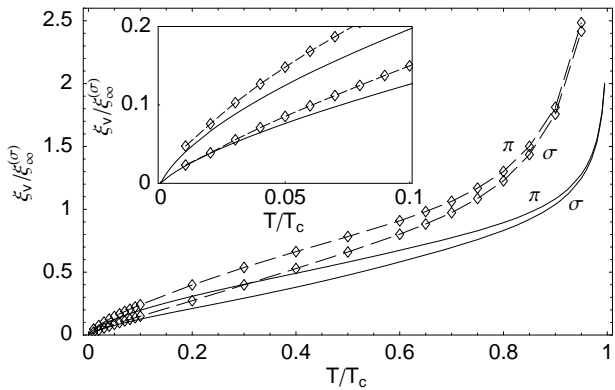}\end{center}

\caption{\label{cap:res2B}The lengthscales $\xi_{v}^{(\sigma)}(T)$ and $\xi_{v}^{(\pi)}(T)$
obtained from a linear profile of the pairing potential (solid line)
and from a numerical self-consistent solution of the gap equation
(diamonds with dashed guidelines). Inset: Blow-up of the low temperature
region.}
\end{figure}

In Fig.~\ref{cap:res2B}, we show the results from these calculations,
again compared to the results from a fully self-consistent solution
of the two band gap equation. As in the single gap case, the linear profile 
for the pairing potential results in a qualitatively correct behaviour of
$\xi_{v}^{(\sigma)}(T)$
and $\xi_{v}^{(\pi)}(T)$ for the full temperature range. The lengthscales
$\xi_{v}^{(\sigma)}$ and $\xi_{v}^{(\pi)}$ diverge for $T\rightarrow T_{c}$
and vanish for $T\rightarrow0$. Thus, we find a Kramer-Pesch effect in both 
bands of a clean two band system. Again, for low temperatures the linear
profile matches nicely the self-consistently calculated curve but
shows deviations for higher temperatures.\\
As in the single band case, the lengthscales $\xi_{v}^{(\sigma)}$
and $\xi_{v}^{(\pi)}$ bend towards $\xi_{v}^{(\sigma)}=\xi_{v}^{(\pi)}=0$
not until very low temperatures ($\sim0.05\, T_{c}$) and thus one
should again be careful with extrapolations from higher temperatures
towards $T=0$ (see inset in Fig.~\ref{cap:res2B}).

\begin{figure}
\begin{center}\includegraphics[%
  width=0.95\columnwidth,
  keepaspectratio]{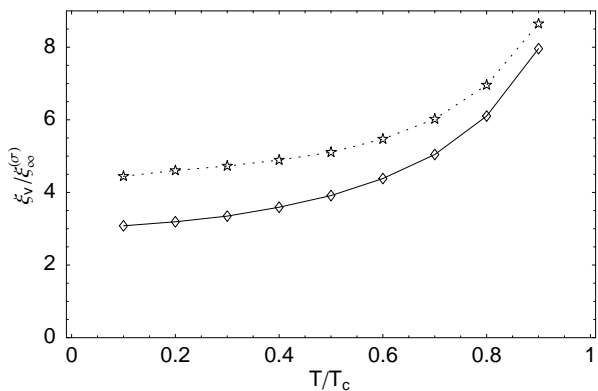}\end{center}

\caption{\label{cap:dirtylim2B}The numerical results for the vortex core radii
$\xi_{v}$ as a function of temperature for the $\sigma$ band (diamonds)
and the $\pi$ band (stars) calculated in the dirty limit with a ratio
of the diffusion constants $\mathcal{D}^{(\sigma)}=0.2\mathcal{D}^{(\pi)}$.
One can clearly see a saturation for low temperatures. The lines are guides for the eye. }
\end{figure}

We also show the comparison with a dirty limit calculation of the
two band gap equation using the Usadel formalism introduced in section
(\ref{sub:dirtylim1}), see Fig.~\ref{cap:dirtylim2B}. In this case
we used slightly different parameters to calculate the coupling matrix
$\lambda_{\alpha\alpha'}$ than in the clean limit calculations. For the
ratio of the diffusion constants we have chosen the value proposed by Koshelev
and Golubov in \cite{Koshelev} as $\mathcal{D}^{(\sigma)}=0.2\mathcal{D}^{(\pi)}$
leading to a ratio of the coherence lengths to be $\xi_{\infty}^{(\sigma)}=0.477\xi_{\infty}^{(\pi)}$.
As expected, we find a finite core size for both bands even for low
temperatures and thus, as in the single
gap case, no Kramer-Pesch effect exists in the dirty limit.

\begin{figure}
\begin{center}\includegraphics[%
  width=0.95\columnwidth,
  keepaspectratio]{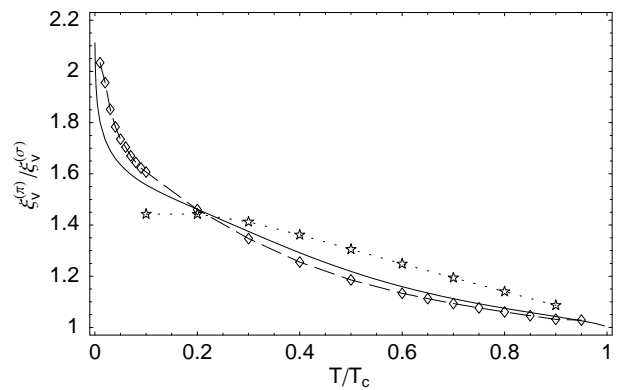}\end{center}

\caption{\label{cap:ratio2B}The ratio $\xi_{v}^{(\pi)}/\xi_{v}^{(\sigma)}$ for the
clean and the dirty limit. Solid line: results from the linear profile of the
pairing potential for the clean limit. Diamonds with dashed guideline:
numerical solution of the gap equation in the clean limit.
Stars with dotted guideline: Usadel formalism results for the dirty limit.}
\end{figure}

In order to compare the behaviour of the two bands, it is useful
to study the ratio $\xi_{v}^{(\pi)}/\xi_{v}^{(\sigma)}$. In
Fig.~\ref{cap:ratio2B}, we show this ratio for the clean and the
dirty limit. The clean limit results were obtained, on the one
hand, from a numerical solution of the gap equation and, on the other
hand, from the analytical model for the pairing potential.
The dirty limit results stem from calculations within the Usadel formalism.
For high temperatures ($T\rightarrow T_{c}$), both $\xi_{v}^{(\pi)}$ and
$\xi_{v}^{(\sigma)}$ diverge and the ratio $\xi_{v}^{(\pi)}/\xi_{v}^{(\sigma)}$
tends to $1$ in the clean as well as in the dirty limit.
For low temperatures ($T\rightarrow 0$), the clean and the dirty limit results 
differ greatly. In the clean limit, both lengthscales
vanish, however, in a way that leads to a strong growth of the
ratio $\xi_{v}^{(\pi)}/\xi_{v}^{(\sigma)}$, i.e. the vortex core size in the
$\pi$ band decreases more slowly as a function of temperature than the one
in the $\sigma$ band. The agreement between the
numerical results and those from the linear profile is very good here.
In the dirty limit, the saturation of both $\xi_{v}^{(\pi)}$ and
$\xi_{v}^{(\sigma)}$ leads to a saturation of the ratio
$\xi_{v}^{(\pi)}/\xi_{v}^{(\sigma)}$.

\begin{figure}
\begin{center}\includegraphics[%
  width=0.8\columnwidth,
  keepaspectratio]{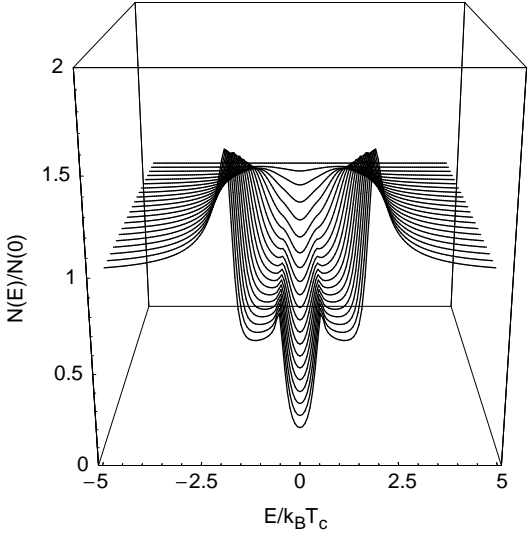}\end{center}

\caption{\label{cap:ldos2Busadelusadel}The quasiparticle spectrum in the
two band dirty limit approach, again for several distances from the
vortex core. As before, the distances span from $r=0$ (rearmost curve)
to $r=10\,\xi_{\infty}$(foremost curve) in steps of $0.5\xi_{\infty}$.}
\end{figure}

The quasiparticle spectrum for two dirty bands is a simple superposition
of two single band spectra as can be seen in Fig.~\ref{cap:ldos2Busadelusadel}.

\section{\label{sec:indkp}Induced Kramer-Pesch Effect in a dirty $\pi$ Band}

\begin{figure}
\begin{center}\includegraphics[%
  width=0.95\columnwidth,
  keepaspectratio]{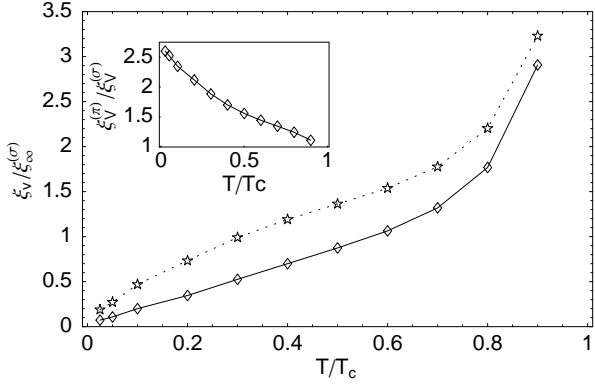}\end{center}

\caption{\label{cap:Mixed_model_results}The numerical results for the votex
core radii $\xi_{v}$ as a function of temperature for the clean $\sigma$
band (diamonds) and the dirty $\pi$ band (stars) calculated in a
mixed model with equal coherence lengths in the two bands 
$\xi^{(\sigma)}_{\infty}=\xi^{(\pi)}_{\infty}$.
The clean $\sigma$ band and the dirty $\pi$ band show both a significant
Kramer-Pesch effect for low temperatures. The ratio of the core radii
in the $\pi$ and in the $\sigma$ band as a function of temperature
calculated for a clean $\sigma$ and a dirty $\pi$ band is shown
in the inset. Again in both plots the lines are guides for the eye. }
\end{figure}

\begin{figure}
\begin{center}
\includegraphics[%
width=0.78\columnwidth,
keepaspectratio]{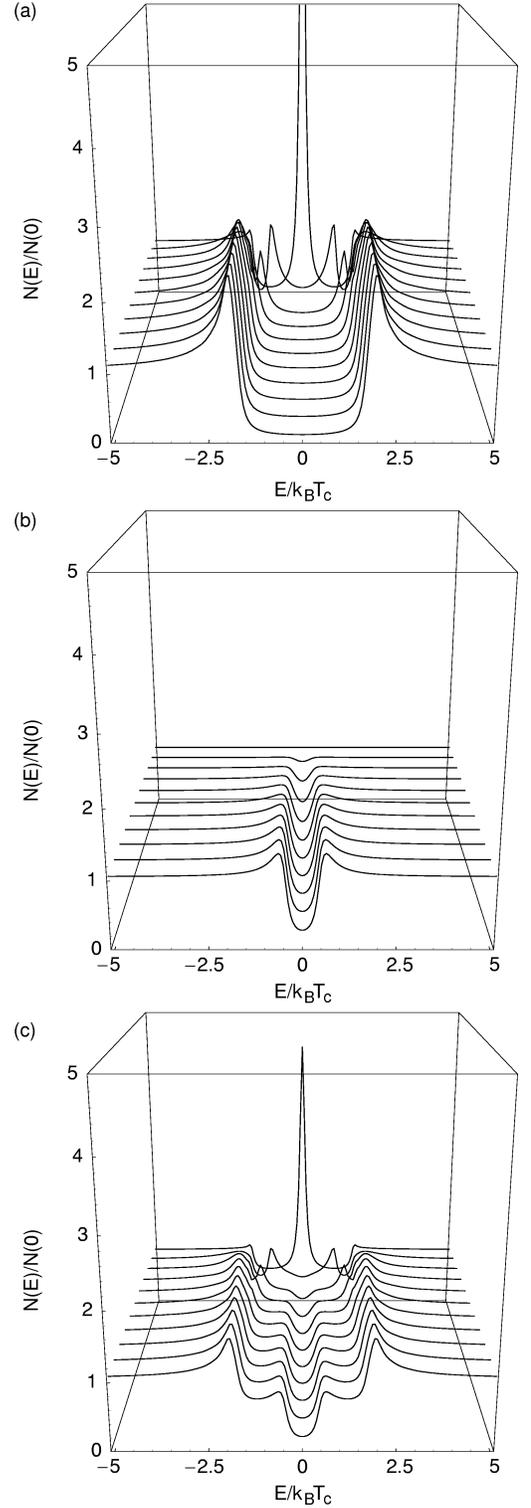}
\end{center}

\caption{\label{cap:ldos2Busadel}The quasiparticle spectrum in the two band
case, again for several distances from the vortex core. Plot (a) shows
the quasiparticle spectrum of the clean $\sigma$ band and (b) that
of the dirty $\pi$ band. In (c), the total quasiparticle spectrum
is being plotted for the mixed model. As before, the distances span
from $r=0$ (rearmost curve) to $r=10\,\xi_{\infty}$(foremost curve)
in steps of $\xi_{\infty}$ and an imaginary part of $\delta=0.1$
was used.}
\end{figure}

In present high-quality samples of MgB$_{2}$ it is believed that the intraband
scattering rate in the $\sigma$ bands is smaller than the gap value,
while in the $\pi$ bands the intraband scattering rate is larger
than the corresponding gap value \cite{Mazin,Quilty,Putti}. 
Therefore we want to discuss in
this section -- as first suggested by M.~Eschrig \cite{Eschrig} -- a model with a clean
$\sigma$ band and a dirty $\pi$ band. Since the gap profiles in
the two bands can not vary independently as they are coupled via the
gap equation, we expect that for low temperatures they
show the same asymptotic behaviour. But it is not at all clear if
we find a decrease of $\xi_{v}^{(\sigma)}$ and $\xi_{v}^{(\pi)}$
for $T\rightarrow0$ as in the clean limit calculations or a saturation
value as in the dirty limit approach. To obtain a result for the two
vortex core radii we have to solve the two band gap equation (\ref{eq:gapeq2B})
where we calculate $f_{\sigma}$ from the Riccati amplitudes $a_{\sigma}$
and $b_{\sigma}$ according to (\ref{eq:propf}), while the Fermi
surface averaged Green's function $\langle f_{\pi}\rangle_{FS,\pi}$ is replaced
by the Usadel Green's function $F_{\pi}(\vec{r},\varepsilon_{n})$ that
is a solution of a simple boundary value problem as described in section
(\ref{sub:dirtylim1}). Since we have two different lengthscales in
the clean and the dirty band case
$\xi^{(\sigma)}_{\infty}=\frac{\hbar v_{F}}{\Delta_{0}^{(\sigma)}}$
and $\xi^{(\pi)}_{\infty}=\sqrt{\frac{\mathcal{D}^{(\pi)}}{2\pi T_{c}}}$ we
have to fix the ratio of the two coherence lengths which in general will depend on the
sample quality, i.e. the scattering rate in the $\pi$ band. 
Assuming the diffusion constant
$\mathcal{D}^{(\pi)}$ to be of the order of 
\[
\mathcal{D}^{(\pi)}\approx2\pi T_{c}\left(\frac{\hbar v_{F}}{\Delta_{0}^{(\sigma)}}\right)^{2}
\]
 we chose the ratio to be $\xi^{(\sigma)}_{\infty}/\xi^{(\pi)}_{\infty}=1$. We do not
expect the essential qualitative results to depend on the special
value of this ratio which we have checked numerically for a different ratio.
In Fig.~\ref{cap:Mixed_model_results} we show
the numerical results of this calculation. In our calculation
we find a distinct decrease of the vortex core size for low temperatures
even in the dirty $\pi$ band, that can be understood as an induced
Kramer-Pesch effect due to the real space coupling of the two gap
amplitudes via the gap equation. The influence of the dirty $\pi$ band
is reflected in a larger ratio of $\xi_{v}^{(\pi)}/\xi_{v}^{(\sigma)}$
for low temperatures than in the pure clean limit or the pure dirty
limit approach.

In Fig.~\ref{cap:ldos2Busadel}, we show the quasiparticle spectrum
for the mixed model, consisting of a clean $\sigma$ band and a dirty
$\pi$ band. Looking at the $\sigma$ and $\pi$ band spectra individually,
it is obvious that the zero energy bound state only exists in the
clean band. Thus, a zero energy peak is visible in the total quasiparticle
spectrum, but it is lifted by an offset due to the flat spectrum
of the dirty band. Restricting to the vortex core region, this leads
to a picture that is very similar to the clean single band case.

\section{\label{sec:concl}Conclusions}

We have studied the Kramer-Pesch effect in both a single gap and a two gap
superconductor using parameters relevant for MgB$_2$. We have presented a model
for the vortex core that allows an analytical solution for the Riccati
equations in both cases and compared it with fully self-consistent numerical
solutions. This model is useful for approximate calculations, in particular
at low temperatures.

In the two gap case we find that the Kramer-Pesch effect is present in both
bands in the clean limit. At high temperatures the sizes of the two
components of the vortex core become equal, while at low temperatures
the size of the vortex core in the $\pi$ band with the smaller gap
decreases more slowly with decreasing temperature.

We also investigated the dirty limit within the framework of the Usadel
equations. In a single band superconductor there is no Kramer-Pesch
effect in the dirty limit as has been discussed before \cite{HayKatSig}.
However, in a two gap superconductor an interesting new situation arises,
when one band is in the clean limit while the other is in the dirty limit,
as is believed to be the case in high quality MgB$_2$ samples.
In this particular case we find that even in the dirty band a Kramer-Pesch 
effect exists, induced by the clean band.

Traditionally the Kramer-Pesch effect is difficult to observe experimentally
due to the disturbing effect of impurities. Our calculations indicate that
there is a better chance of observing this effect in MgB$_2$ by scanning
tunneling microscopy (STM) imaging of the vortex core as a function of
temperature, because the $\pi$ band may be in the dirty limit and still
showing a decreasing vortex core size with decreasing temperature, as
seen in Fig.~\ref{cap:Mixed_model_results}.

\acknowledgments

We would like to thank Christian Iniotakis, Nobuhiko Hayashi and Matthias Eschrig for valuable
discussions. T.~Dahm acknowledges funding from a research grant for young scientists
of the University of T\"ubingen.

\end{document}